\documentclass[english]{revtex4}
\usepackage[T1]{fontenc}
\usepackage[latin9]{inputenc}
\usepackage{graphicx}

\makeatletter

\makeatletter

\usepackage{hyperref}

\usepackage{bm}

\makeatother

\usepackage{babel}
\makeatother

\begin{document}

\title{Stereo-photography of point-plane streamers in air}

\author{S~Nijdam\footnote{Corresponding e-mail: s.nijdam@tue.nl}, J~S
Moerman, T~M~P Briels, E~M van Veldhuizen}

\affiliation{Eindhoven University of Technology, Dept.\ Applied Physics\\
 P.O. Box 513, 5600 MB Eindhoven, The Netherlands}

\author{U Ebert}

\affiliation{Centrum Wiskunde \& Informatica, Amsterdam, and Eindhoven Univ.\ Techn.,
The Netherlands}

\begin{abstract}
Standard photographs of streamer discharges show a two-dimensional
projection. We here present stereo-photographic images that resolve
their three-dimensional structure. We describe the stereoscopic set-up
and evaluation, and we present results for positive streamer discharges
in air at 0.2~-~1~bar in a point plane geometry with a gap distance
of 14~cm and a voltage pulse of 47~kV. In this case an approximately
Gaussian distribution of branching angles of $43\textdegree\pm12\textdegree$
is found; these angles do not significantly depend on the distance
from the needle or on the gas pressure.
\end{abstract}
\maketitle
A streamer is a rapidly extending discharge channel that can appear
when a high voltage is applied to any ionizable medium; most studies
are done in air. Streamers precede phenomena like sparks, leaders
and lightning. The main difference is that streamers do not significantly
increase the gas temperature; they are rather governed by impact ionization
and space charge effects~\citep{Ebert2006}. Streamers are directly
observed in nature in the form of sprites~\citep{Pasko2007}, that
are enormous atmospheric discharges above active thunderstorms at
about 40~-~90~km altitude. Streamers also have many technical applications,
in ozone generation and consecutive disinfection, in bio fuel processing,
plasma assisted combustion and aviation; for a short review with references,
we refer to~\citep{Ebert2006}.

A largely unexplored issue in streamer research is the breakup of
single channels into many. Such branching events are commonly seen
in experiments~\citep{Veldhuizen2002,Briels2006,Briels2008a}; multiple
branching actually determines the gas volume that is crossed by streamers
and consecutively chemically activated for plasma processing purposes.
However, up to now, only the conditions of the first branching event
have been resolved in microscopic models \citep{Liu2004,Arrayas2002,Montijn2006,Luque2007,Pancheshnyi2005}.
On the other hand, the distribution of branching lengths and angles
is an ingredient of models for the complete branching tree on larger
scales \citep{Niemeyer1984,Akyuz2003,Pasko2001}. In the present paper,
we resolve these lengths and, in particular, the angles in experiments.

Imaging of streamer discharges is usually done with conventional or
digital cameras \citep{Pancheshnyi2005,Briels2006,Winands2006}. This
leads to two-dimensional (2D) representations of what is essentially
a three-dimensional (3D) phenomenon. These 2D representations can
cause problems of interpretation. For example, it is impossible to
see whether an apparent loop or reconnection is really what it seems
to be. It is also impossible to get a complete picture of the 3D spatial
structure and to measure branching angles. For this purpose, we have
implemented a stereo-photography method which makes it possible to
image streamer discharges in 3D. In this way, we resolve the imaging
ambiguities in the fundamental physical phenomena, help understanding
which gas volumes are actually treated by the discharge, and supply
experimental data for larger scale models. The stereoscopic technique
that we use has been around for a very long time \citep{Brewster1856,Faugeras1993}
and has been used for a large variety of topics. Some phenomena similar
to streamers that have been studied with stereo-photography are sparks
\citep{MacAlpine1999}, flames \citep{Ng2003} and dusty plasmas \citep{Jr2004}.

\begin{figure}[h]
\includegraphics[width=9cm]{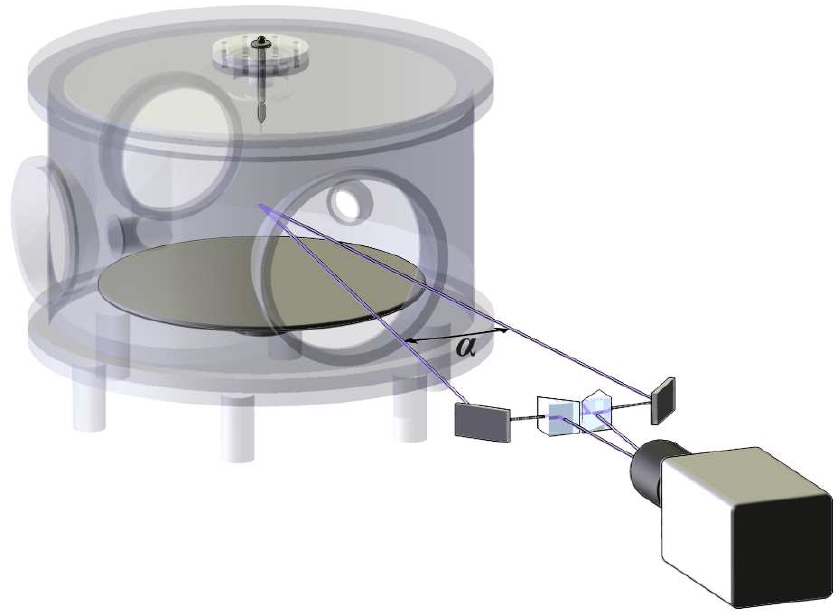}\caption{\label{fig:Setup}Overview of the stereoscopic measurement set-up
with a schematic drawing of the 2 image paths.}

\end{figure}

To generate streamers, we use the experimental set-up that is discussed
thoroughly in~\citep{Briels2006}, and we use the electric circuit
called C-supply in~\citep{Briels2006}. In this set-up a capacitor
is charged negatively with a DC power supply. This capacitor is then
discharged by means of a spark-gap switch. This results in a positive
voltage peak on the needle inside the vacuum vessel. A positive corona
discharge then propagates from the needle to the grounded plate. Both
needle and plate are highlighted in figure~\ref{fig:Setup}. In the
present measurements, a positive voltage of 47~kV with a rise-time
of about 30~ns was applied to the point, 14~cm above the plate.
The atmosphere in the vacuum vessel consisted of ambient air at different
pressures (200, 565 and 1000~mbar).

MacAlpine et al. \citep{MacAlpine1999} have studied sparks with a
camera and a prism. In this study two images were taken using a prism
to form an image at a right angle to the directly-observed one. In
this way the complete 3D-structure of the spark path can be reconstructed
with great accuracy. Similar work was reported by M.\ Makarov \citep{Makarov}.
However, this method only works well for structures that have very
few channels (e.g. the one spark of \citep{MacAlpine1999}). When
there are many channels, it is very difficult to correlate them pairwise
from two images taken at an angle of 90$\textdegree$.

\begin{figure}[t]
\includegraphics[width=9cm]{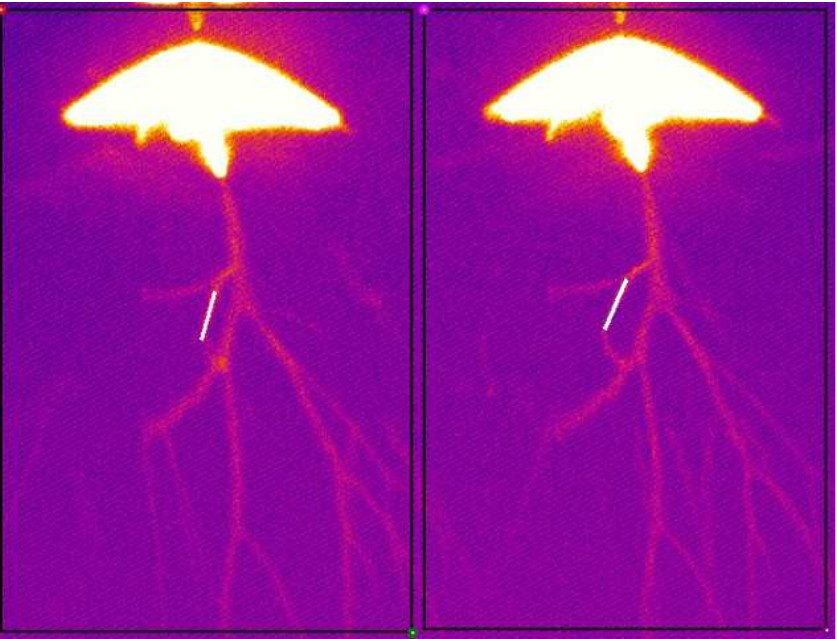}\caption{\label{fig:Stereo-image}Stereo image as recorded by camera. Settings:
positive voltage on tip, \textit{$U=47\mathrm{kV}$}, $p=200\mathrm{mbar}$,
$\alpha=13\textdegree$, $d=14\mathrm{cm}$. The intensity has been
scaled so that the structure in the bottom part can be clearly seen.
One streamer section has been marked with a white line in both images.}

\end{figure}
In our case, we want to study streamer discharges that contain many
(10-100) streamers. For this purpose a similar method can be used,
but with a much smaller angle between the two image paths so that
the two images of one streamer can be recognized. To achieve a smaller
angle, one camera has been used in combination with two prisms and
two flat mirrors as shown in figure~\ref{fig:Setup}. With this set-up
two images (from different viewing angles) are captured on one camera
frame; therefore they are temporarily perfectly synchronized. An example
of such a camera frame is shown in figure~\ref{fig:Stereo-image}.

\begin{figure}
\includegraphics[width=9cm]{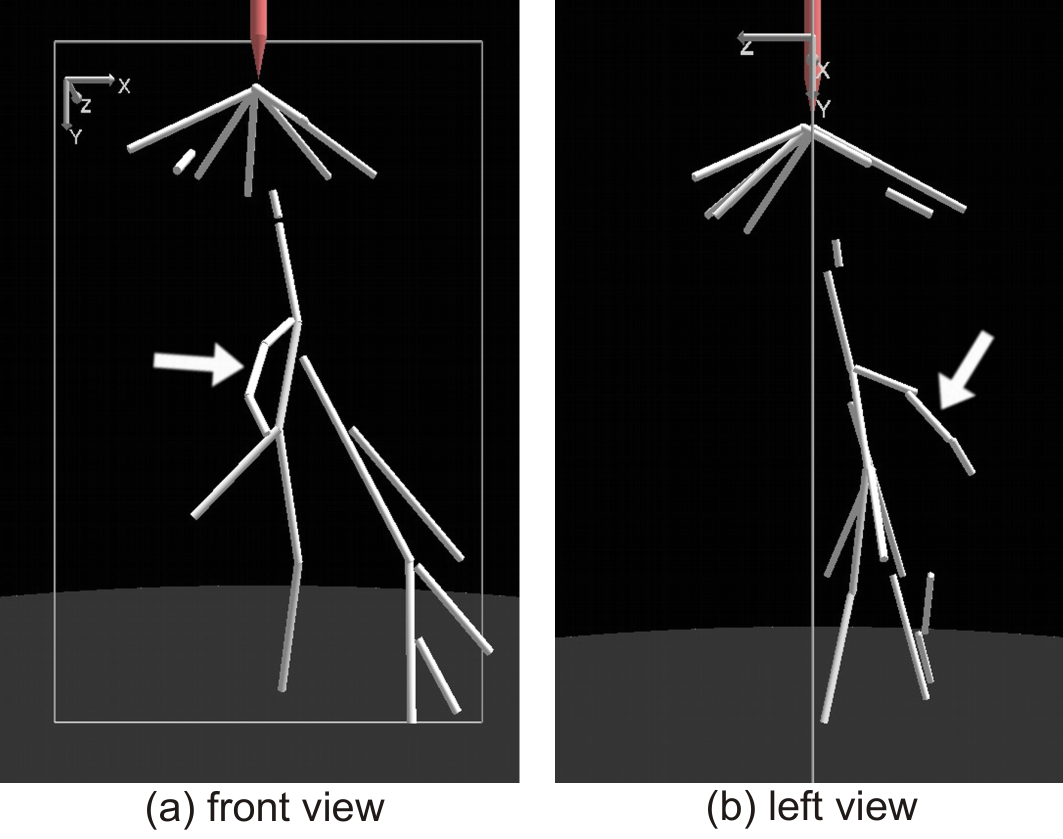}\caption{\label{fig:3D-reconstruction}Orthogonal views of the 3D reconstruction
of streamer structure shown in figure~\ref{fig:Stereo-image}. The
section originally marked with the white line is now marked with an
arrow in both views.}

\end{figure}
From the two 2D-images the 3D structure of the streamer channels can
be reconstructed in the following manner: a straight section of a
streamer channel is selected in both images. The end points of these
two lines are now translated from 2D (xy) to 3D (xyz). In principle,
an exact trigonometric evaluation would supply absolute locations
in space. However, as we are only interested in local observables
(branching angles and lengths), we have used a simplified approach
assuming that the cameras are far from the system and have a very
large focal length. Indeed, the distance between camera and streamers
is about 1~m, while distances between recently splitted streamer
branches never exceed 2 cm.

The two images give the 2D coordinates $(x_{l},y_{l})$ and $(x_{r},y_{r})$
of identical streamer parts within the left or right image respectively,
where the origins of the respective coordinate systems are chosen
in the electrode tip. The depth coordinate $z$ is then approximated
as $z=\left(x_{r}-x_{l}\right)/\left(2\cdot\sin\left(\alpha/2\right)\right)$,
where $\alpha$ is the full angle between the two optical paths (as
indicated in figure~\ref{fig:Setup}, in the present measurements
$\alpha=13\textdegree$). The 3D $x$ and $y$-coordinates are calculated
as $x=\left(x_{r}+x_{l}\right)/2$ and $y=\left(y_{r}+y_{l}\right)/2$.

The error in streamer distances after splitting that results from
this simplification is less than 0.2~mm. The dominant error comes
from the visual determination of the locations of streamer section
end points on the stereoscopic images. In many situations it is difficult
to locate the exact point of branching, especially where two streamers
are very close to each other. The total error is approximately 1~mm
for local observables and 5~mm for absolute locations.

The two 2D lines have now been translated into one 3D streamer section.
This can be done for all suitable streamer sections in the image.
When all these 3D streamer sections are now plotted in 3D-space, we
get some insight in the real structure of the streamer discharge.
The 3D reconstruction of the example from figure~\ref{fig:Stereo-image}
is shown in figure~\ref{fig:3D-reconstruction}. Here it can clearly
be seen that the streamer section marked with the white lines in figure~\ref{fig:Stereo-image}
is not part of a loop. This information can not be derived from just
one of the original 2D images. One of the measurements that can be
performed now is measuring branching angles. The measured angles are
the inner angles between two 3D streamer sections, represented as
vectors. The technique described here also has some limitations, the
most important one is that it is not possible to process discharge
images that contain more than about 50 streamer channels.

\begin{figure*}[t]
\includegraphics{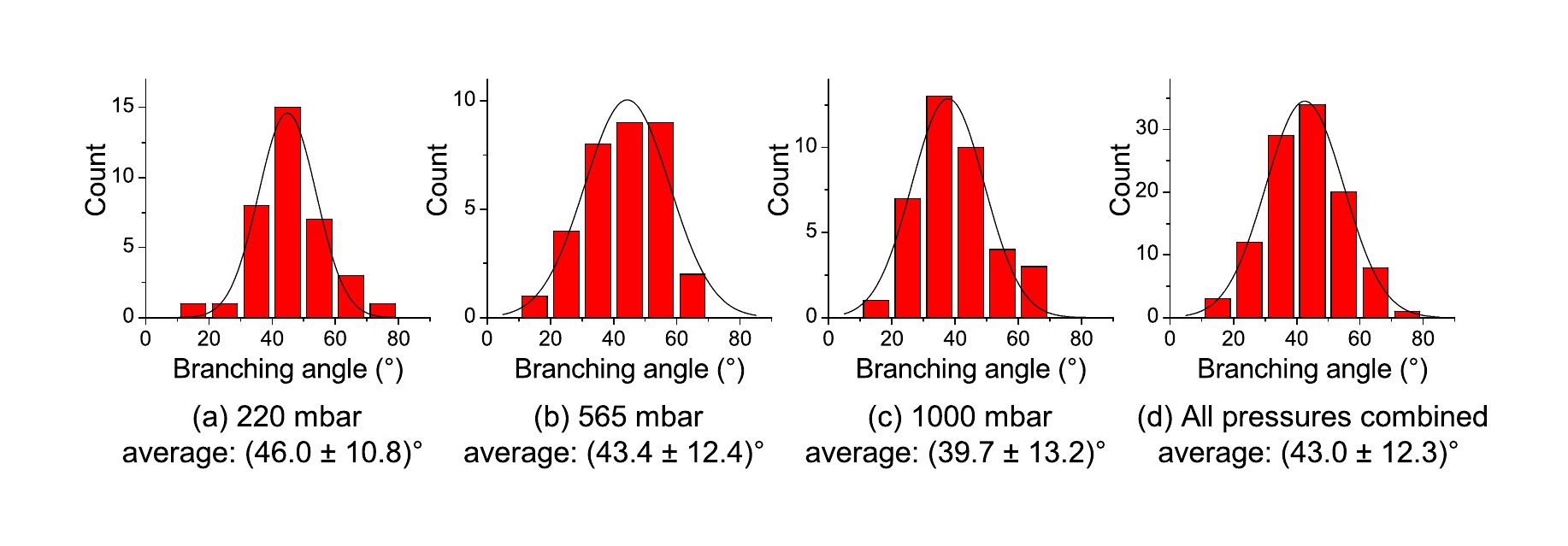}\caption{\label{fig:histograms}Histograms with Gaussian fits for branching
angles for three different pressures and for all pressures combined.}

\end{figure*}
Figures~\ref{fig:histograms}a-c show histograms of the measured
branching angles for 200, 565 and 1000~mbar and figure~\ref{fig:histograms}d
combines the results for all pressures into one histogram. As can
be seen, the distribution is roughly Gaussian, with average values
between 39$\textdegree$ and 46$\textdegree$ and standard deviations
of 11$\textdegree$ to 13$\textdegree$. The average branching angle
shows a slight decrease as a function of pressure. However, it is
not clear whether this is statistically significant due to the limited
amount of data points (about 35 points per pressure setting).

\begin{figure}
\includegraphics{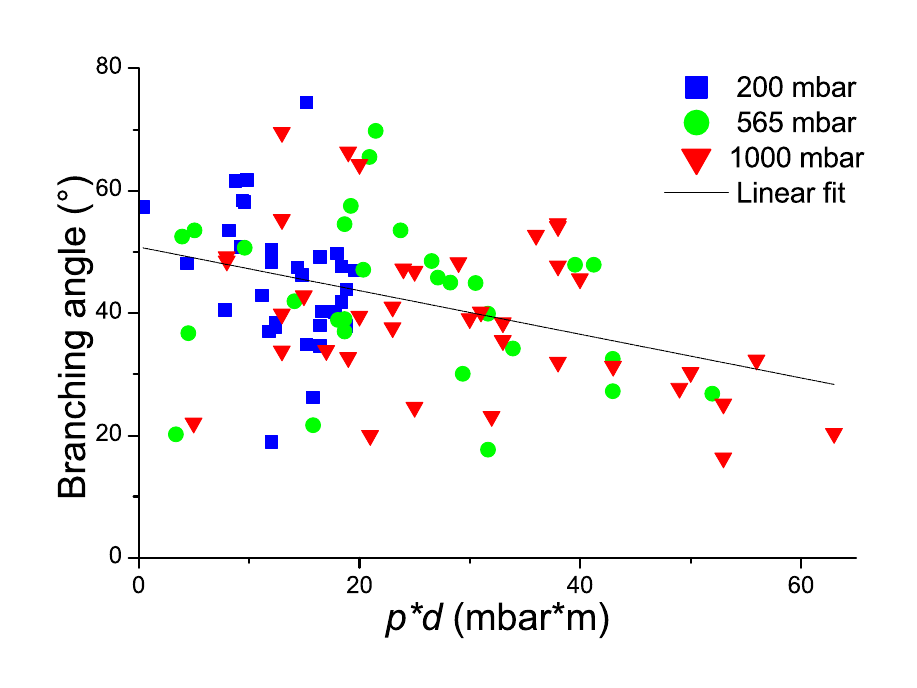}\caption{\label{fig:anglevsposition}Measured branching angle as function of
$p\cdot d$.}

\end{figure}
The length scales of streamers are expected and observed to scale
quite well with pressure. However, density fluctuations do not scale
with density \citep{Ebert2006,Briels2008}; if they play a significant
role in streamer branching, one would expect the branching distribution
to depend on pressure. Therefore, in figure~\ref{fig:anglevsposition}
the branching angle is plotted as function of $p\cdot d$ where $p$
is the pressure and $d$ the vertical distance from the tip (the $y$-coordinate)
at the point of branching. If the branching behaviour would differ
for streamer sections close to the tip from sections close to the
cathode plane, this would be visible in this plot. Also a pressure
dependence would be visible. However, only a small dependence on $p\cdot d$
can be observed. This dependence is statistically not significant
given the large spread and measurement error in the data set (correlation
coefficient $R^{2}=0.15$).

The ratio of streamer length between branching events over streamer
width has also been measured. This ratio is about 15 for all pressures.
This is a bit higher than the ratio of 12 found by Briels \textit{et
al.}~\citep{Briels2008}.

In conclusion, we have built a stereographic set-up that is able to
reconstruct 3D spatial structures of streamer discharges. This enables
us to get more insight into what really happens in such a discharge.
For example, we are now able to see if something that looks like a
streamer reconnecting to another streamer is indeed what it seems.
Up to now, such statements relied on multiple observations from 2D
images \citep{Briels2006}. We are also able to measure branching
angles of streamers.

We have found that the branching angle for streamers in an overvolted
gap of 16 cm does not significantly depend on pressure and $p\cdot d$
and is distributed normally with an average of 43$\textdegree$ and
a standard deviation of 12$\textdegree$.

S.N. and T.B. acknowledge support by STW-project 06501, part of the
Netherlands Organisation for Scientific Research NWO.

\end{document}